\input{aipcheck}

\documentclass[draft
    ,final            
  ]
  {aipproc}

\layoutstyle{6x9}

\begin{document}

\title{$0^{++}$ Resonances Observed at BES}


\author{B. S. Zou (for BES Collaboration)}{
  address={Institute of High Energy
Physics, CAS, Beijing 100049, China} }

\begin{abstract}
In last 10 years, $0^{++}$ resonances have been observed and studied
at BES in many processes, such as
$J/\psi\to\gamma\pi^+\pi^-\pi^+\pi^-$, $\gamma\pi^+\pi^-$,
$\gamma\pi^0\pi^0$, $\gamma K^+K^-$, $\gamma K_SK_S$,
$\gamma\omega\phi$, $\omega\pi^+\pi^-$, $\omega K^+K^-$,
$\phi\pi^+\pi^-$, $\phi K^+K^-$, $\psi(2S)\to J/\psi\pi^+\pi^-$,
$\chi_{c0}\to\pi^+\pi^-K^+K^-$, $\pi^+\pi^-\pi^+\pi^-$ etc.. The
results on $0^{++}$ resonances observed at BES are reviewed.
\end{abstract}

\maketitle


\section{Introduction}

The study of the isoscalar $0^{++}$ resonances is of crucial
importance for understanding the whole hadron spectrum. The lightest
$q\bar q$ spatial excited state, the lightest glueball, and the
lightest tetra-quark state, all are expected to have such quantum
numbers of vacuum. Up to now, there are five isoscalar $0^{++}$
resonances ($\sigma/f_0(600)$, $f_0(980)$, $f_0(1370)$, $f_0(1500)$
and $f_0(1710)$) listed as well-established ones by PDG \cite{PDG},
and four unestablished ones ($f_0(1790)$, $f_0(2020)$, $f_0(2100)$
and $f_0(2200)$). None of them gets a clear picture about its
internal structure. They are ascribed as $q\bar q$ states, $q^2\bar
q^2$ states, meson-meson molecules. glueballs, coupled-channel
dynamical states, etc.

Among various reactions, three kinds of processes from charmonium
decays may play important role for understanding the nature of these
scalars and have been studied by BES Collaboration at Beijing
Electron-Positron Collider (BEPC). They are $\psi$ radiative decays,
$\psi$ hadronic decays against $\phi$ or $\omega$, and $\chi_{c0}$
decays. In the following three sections, we will outline the major
physics roles and review main results on scalar resonances from each
of them. A brief summary of the information on scalars we deduced
from these processes is given in the final section.

\section{$0^{++}$ resonances observed in $\psi$ radiative decays}

There are three main physics objectives for {$\psi$ radiative
decays:

(1) Looking for glueballs and hybrids. As shown in Fig.~\ref{fig:1},
after emitting a photon, the $c\bar c$ pair is in a $C=+1$ state and
decays to hadrons dominantly through two gluon intermediate states.
Simply counting the power of $\alpha_s$ we know that glueballs
should have the largest production rate, hybrids the second, then
the ordinary $q\bar q$ mesons.

(2) Completing $q\bar q$ meson spectroscopy and studying their
production and decay rates, which is crucial for understanding their
internal structure and confinement.

(3) Extracting $gg\leftrightarrow q\bar q$ coupling from
perturbative energy region of above 3.6 GeV to nonperturbative
region of 0.3 GeV. This may show us some phenomenological pattern
for the smooth transition from perturbative QCD to strong
nonperturbative QCD.

\begin{figure}[htbp]
\includegraphics[width=14cm,height=5cm]{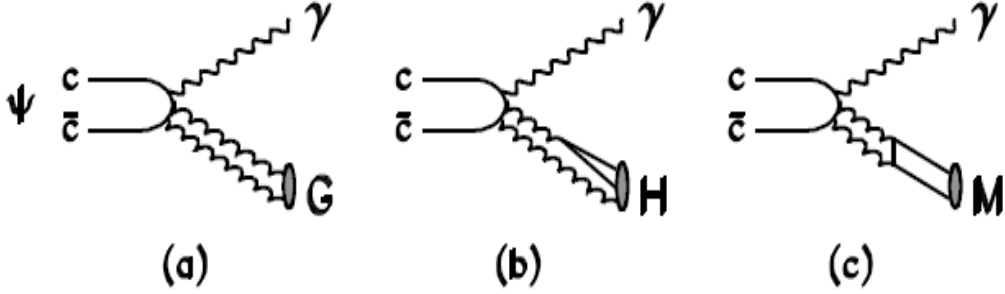}
\caption{$\psi$ radiative decays to (a) glueball, (b) hybrid, and
(c) $q\bar q$ meson. } \label{fig:1}
\end{figure}

For the study of the isoscalar $0^{++}$ resonances from the $J/\psi$
radiative decays, we have studied the largest radiative decay
channel $J/\psi\to\gamma 4\pi$, the simplest ones
$J/\psi\to\gamma\pi\pi$, $\gamma K\bar K$, and the exotic doubly OZI
suppressed one $J/\psi\to\gamma\omega\phi$. Partial wave analysis
has been performed for the $\gamma\pi^+\pi^-\pi^+\pi^-$ channel of
BESI data~\cite{BESg4pi}, $\gamma\pi^+\pi^-$~\cite{BESg2pi}, $\gamma
K^+K^-$~\cite{BESg2K} and $\gamma\omega\phi$~\cite{BESgomegaphi}
channels of BESII data.

The invariant mass spectra and $0^{++}$ partial wave contribution
for $J/\psi\to\gamma\pi^+\pi^-\pi^+\pi^-$ at BESI~\cite{BESg4pi} and
$\gamma\pi^+\pi^-$ at BESII~\cite{BESg2pi} are shown in
Fig.\ref{fig:g4pi-g2pi}. The two channels have a similar pattern of
three-peak structure for the $0^{++}$ partial wave contribution. The
fitted mass and width for these $0^{++}$ peaks in
$\gamma\pi^+\pi^-\pi^+\pi^-$~\cite{BESg4pi} and
$\gamma\pi^+\pi^-$~\cite{BESg2pi} channels are listed in
Table~\ref{tab:a}. The results from the two channels are consistent
with each other within their error bars and with those from a
reanalysis of MARKIII data for the $\gamma\pi^+\pi^-\pi^+\pi^-$
channel~\cite{Mark3}. The $2\pi$ invariant mass spectrum for the
$\gamma\pi^0\pi^0$ channel is similar to that from the
$\gamma\pi^+\pi^-$ channel as shown in Fig.~\ref{fig:gKK-g2pi}.

\begin{figure}[htbp]
\includegraphics[width=15cm,height=6cm]{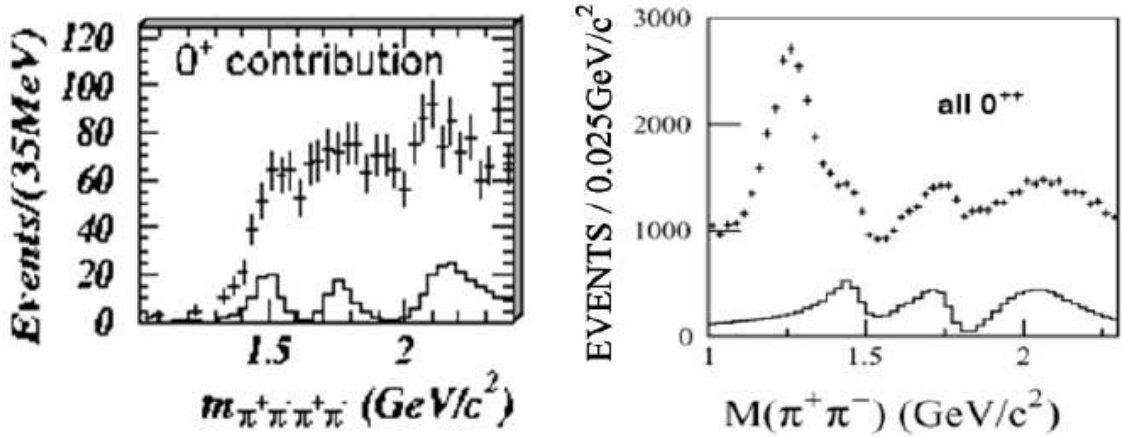}
\caption{Invariant mass spectra and $0^{++}$ partial wave
contribution for $J/\psi\to\gamma\pi^+\pi^-\pi^+\pi^-$ at
BESI~\cite{BESg4pi} (left) and $\gamma\pi^+\pi^-$ at
BESII~\cite{BESg2pi} (right) } \label{fig:g4pi-g2pi}
\end{figure}

\begin{table}
\begin{tabular}{lcccc}
\hline resonance & mass~\cite{BESg4pi}
  & width~\cite{BESg4pi}
  & mass~\cite{BESg2pi}
  & width~\cite{BESg2pi}   \\
\hline
$f_0(1500)$ & $1505^{+15}_{-20}$ & $140^{+40}_{-30}$ & $1466\pm 6\pm 16$ & $108^{+14}_{-11}\pm 21$\\
$f_0(1710\sim 1790)$ & $1740^{+30}_{-25}$ & $120^{+50}_{-40}$ & $1765^{+4}_{-3}\pm 11$  & $145\pm 8\pm 23$\\
$f_0(2020\sim 2200)$ & $2090\pm 30$ & $330\pm 100$ & 2020 PDG  & PDG\\
\hline
\end{tabular}
\caption{Fitted mass and width (in unit of MeV) for the three
$0^{++}$ peaks in $\gamma\pi^+\pi^-\pi^+\pi^-$~\cite{BESg4pi} and
$\gamma\pi^+\pi^-$~\cite{BESg2pi} channels} \label{tab:a}
\end{table}

The invariant mass spectra for the $J/\psi\to\gamma K^+K^-$ and
$\gamma K_S K_S$ channels are shown in Fig.~\ref{fig:gKK-g2pi}. The
partial wave analysis has been performed for the mass below 2
GeV~\cite{BESg2K}. While the peak around 1525 MeV is mainly due to
$f_2^\prime(1525)$, the peak around 1740 MeV is dominated by
$0^{++}$ contribution with fitted mass and width as $1740\pm
4^{+10}_{-25}$ and $166^{+5+15}_{-8-10}$ MeV, respectively.
Recently, the $\gamma\pi^+\pi^-$ and $\gamma K^+ K^-$ channels from
$\psi^\prime$ decays have also been studied with similar peaks
observed~\cite{BESg2pi(2S)}.

\bigskip

\begin{figure}[htbp]
\includegraphics[width=14cm,height=10cm]{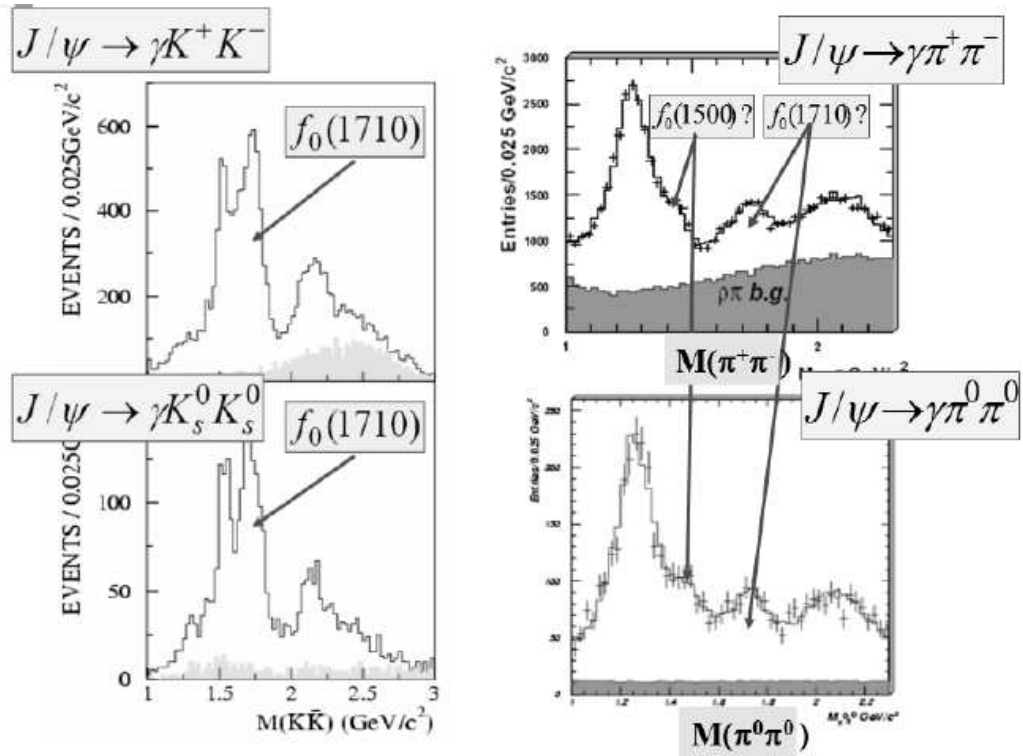}
\caption{Invariant mass spectra for the $J/\psi\to\gamma K\bar
K$~\cite{BESg2K} and $\gamma\pi\pi$~\cite{BESg2pi} channels}
\label{fig:gKK-g2pi}
\end{figure}

\begin{figure}[htbp]
\includegraphics[width=14cm,height=6cm]{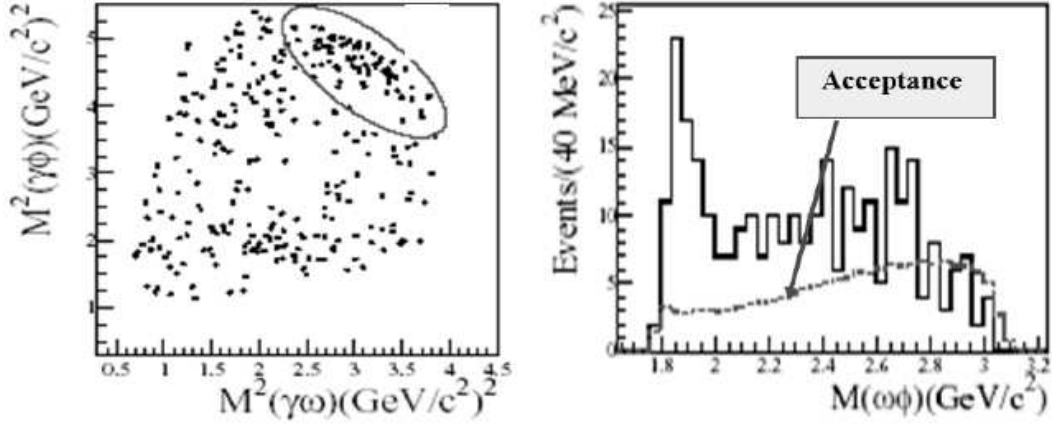}
\caption{Dalitz plot and Invariant mass spectrum for the
$J/\psi\to\gamma\omega\phi$ channel~\cite{BESgomegaphi}.}
\label{fig:gphi-omega}
\end{figure}

For the doubly OZI suppressed channel $\gamma\omega\phi$, the Dalitz
plot and the $\omega\phi$ invariant mass spectrum are shown in
Fig.~\ref{fig:gphi-omega}. A clear enhancement near $\omega\phi$
threshold is observed. A partial wave analysis shows that this
enhancement favors $J^P=0^+$~\cite{BESgomegaphi}. If fitted with a
simple Breit-wigner formulae of constant width, the mass and width
are obtained as $M=1812^{+19}_{-26}\pm 18$ MeV and $\Gamma=105\pm
20\pm 28$ MeV, respectively. Considering possible energy dependence
factors caused by threshold and off-shell effects, the possible
relation with the sub-threshold $f_0(1710\!\sim\!1790)$ structure
cannot be excluded. But the large branching ratio to this doubly OZI
suppressed channel needs some special theoretical attention to
identify its nature~\cite{X1810}.

\section{$0^{++}$ resonances observed in $\psi$ hadronic decays }

For the $\psi$ hadronic decays against $\phi/\omega$, there are also
mainly three physics objectives:

(1) Looking for hybrids. Since $\psi$ decays to hadrons through
three gluons, final states involving a hybrid as shown in
Fig.~\ref{fig:2}(a) are expected to have larger production rate than
ordinary $q\bar q$ mesons as shown in Fig.~\ref{fig:2}(b,c).

(2) Extracting $u\bar u+d\bar d$ and $s\bar s$ components of
associated mesons, M, via $\psi\to M+\omega/\phi$ as shown in
Fig.~\ref{fig:2}(b,c).

(3) Disfavoring glueball production. We can analyze the quark/gluon
content of a particle, M, by comparing its production in
$\psi\to\gamma M$, $\omega M$ and $\phi M$.

\begin{figure}[htbp]
\includegraphics[width=14cm,height=5cm]{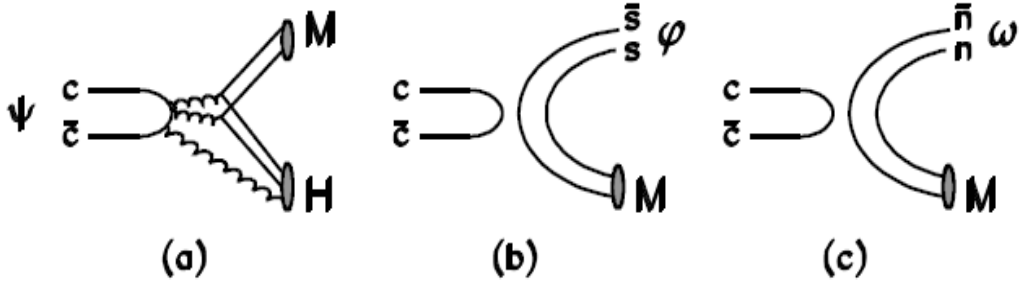}
\caption{$\psi$ hadronic decays to (a) hybrids, (b) $s\bar s$, and
(c) $n\bar n\equiv {1\over\sqrt{2}}(u\bar u+d\bar d)$ mesons. }
\label{fig:2}
\end{figure}

To investigate the $u\bar u+d\bar d$ and $s\bar s$ components of
iso-scalar $0^{++}$ resoances, we have studied
$J/\psi\to\omega\pi^+\pi^-$~\cite{BESomega2pi}, $\omega
K^+K^-$~\cite{BESomega2K}, $\phi\pi^+\pi^-$ and $\phi
K^+K^-$~\cite{BESphi2pi} channels. The invariant mass spectra for
these channels are shown in Fig.\ref{fig:phiKK}. They are similar to
the previous ones by MARKIII and DM2 Collaborations, but with much
higher statistics.

\begin{figure}[htbp]
\includegraphics[width=14cm,height=10cm]{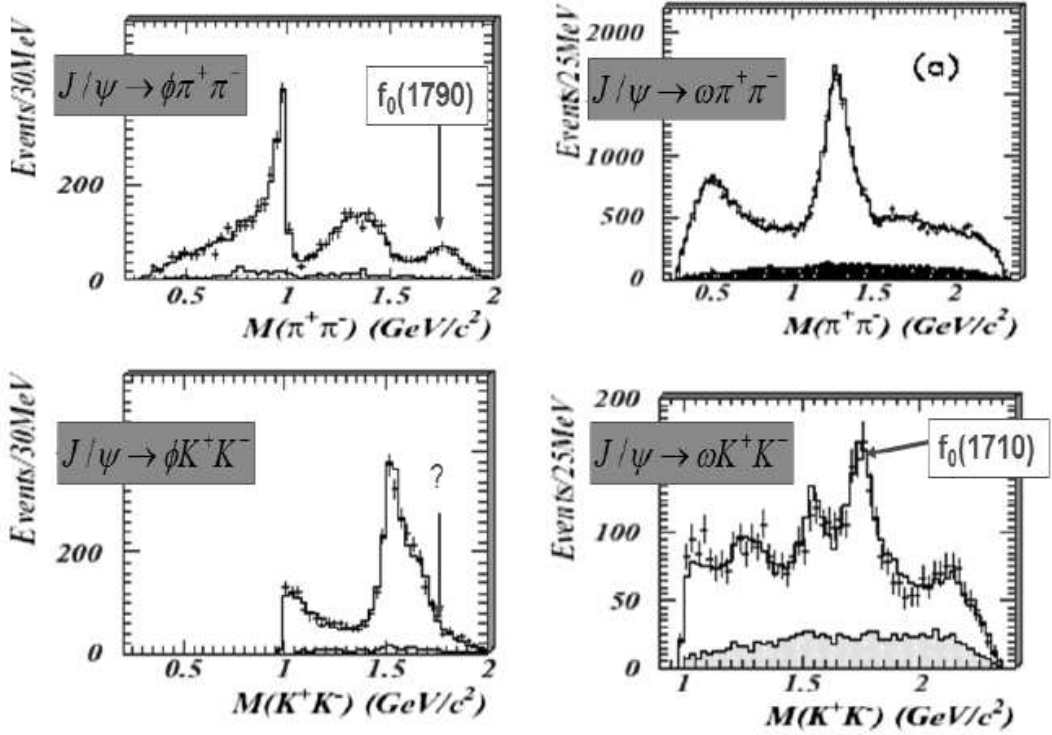}
\caption{Invariant mass spectra for the $J/\psi\to\phi\pi^+\pi^-$,
$\phi K^+ K^-$~\cite{BESphi2pi},
$\omega\pi^+\pi^-$~\cite{BESomega2pi} and $\omega K^+
K^-$~\cite{BESomega2K} channels.} \label{fig:phiKK}
\end{figure}

For $J/\psi\to\omega\pi^+\pi^-$, there are two clear peaks at 500
MeV and 1275 MeV in the $2\pi$ mass spectrum corresponding to the
$\sigma$ and the $f_2(1275)$, respectively. The $\sigma$ pole
position is determined to be $(541 \pm 39)-i(252 \pm 42)$ MeV from
the mean of six analyses~\cite{BESomega2pi} by using various
parametrizations~\cite{zou1,zheng}. It is also consistent with a
later study of high statistical data of $\psi^\prime\to
J/\psi\pi^+\pi^-$~\cite{BESpsi2pi}. For $J/\psi\to\omega K^+K^-$,
there is a conspicuous signal for $f_0(1710)\to K^+ K^-$. From a
combined analysis of these two sets of data, the branching ratio
$BR(f_0(1710)\to\pi\pi)/BR(f_0(1710)\to K\bar K)$ is $<0.11$ at the
95\% confidence level~\cite{BESomega2K}.

For the $J/\psi\to\phi\pi^+\pi^-$ and $\phi K^+K^-$ channels, the
$f_0(980)$ is observed clearly in both sets of data, and parameters
of the Flatt\' e formula are determined accurately: $M = 0.965\pm
0.010$ GeV, $g_1 = 0.165 \pm 0.018 $ GeV$^2$, $g_2/g_1 = 4.21 \pm
0.25 \pm 0.21$. The $\phi \pi \pi$ data also exhibit a strong $\pi
\pi$ peak centered at $M = 1335$ MeV. It may be fitted with
$f_2(1270)$ and a dominant $0^+$ signal made from $f_0(1370)$
interfering with a smaller $f_0(1500)$ component. There is also a
state in $\pi \pi$ with $M = 1790 ^{+40}_{-30}$ MeV and $\Gamma =
270 ^{+60}_{-30}$ MeV; spin 0 is preferred over spin 2. A particular
feature is that $f_0(1790)\to\pi\pi$ is strong, but there is little
or no corresponding signal for decays to $K\bar K$. This behavior is
incompatible with $f_0(1710)$, which is known to decay dominantly to
$K\bar K$. So this state, $f_0(1790)$, is distinct from $f_0(1710)$.
The $\phi K\bar K$ data contain a strong peak due to $f_2'(1525)$. A
shoulder on its upper side may be fitted by interference between
$f_0(1500)$ and $f_0(1710)$.

The large production rates of $f_2(1275)$ against $\omega$ and
$f'_2(1525)$ against $\phi$ demonstrate the flavor filter role of
the $\psi$ hadronic decays against $\phi/\omega$. For the scalar
production, the $\omega\sigma/f_0(600)$ and $\phi f_0(980)$ have the
largest branching ratios. This suggests that the $f_0(980)$ has
large strangeness content while the $\sigma/f_0(600)$ contains
mainly the non-strange quarks, just as expected by various
models~\cite{f0(980)}. It seems indeed that the non-glueballs are
favorably produced in these reactions. While $\phi f_0(1370)$, $\phi
f_0(1790)$ and $\omega f_0(1710)$ have also significant production
rate, the $f_0(1500)$ is hardly visible in any of these processes.

\section{$0^{++}$ resonances observed in $\chi_{c0}$ decays}

Since the decays of $\chi_{c0}$ and $\chi_{c2}$ into light hadrons
are mainly through intermediate states of two gluons, they are
expected to provide the cleanest place for studying
gluon-hadronization dynamics. Schematic pictures for the decays of
$\chi_{c0,2}$ into meson pairs via the production of different
components are shown in Fig.\ref{fig:chi-decay}. The study of these
decays may also shed light on the internal structures of the
produced mesons.

\begin{figure}[htbp]
\includegraphics[width=14cm,height=3cm]{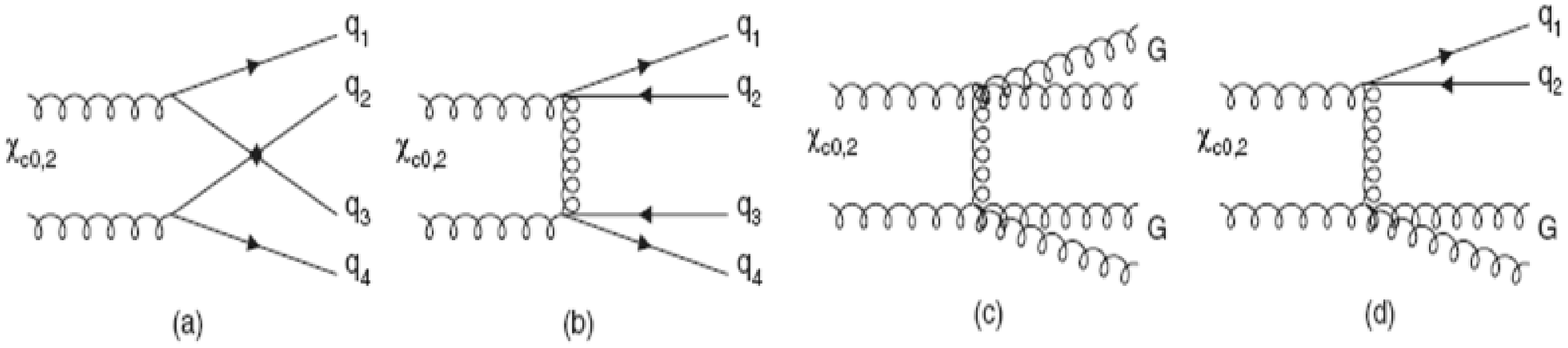}
\caption{Schematic pictures for the decays of $\chi_{c0,2}$ into
meson pairs via the production of different components~\cite{zhao2}.
} \label{fig:chi-decay}
\end{figure}

\bigskip

\begin{figure}[htbp]
\includegraphics[width=14cm,height=6cm]{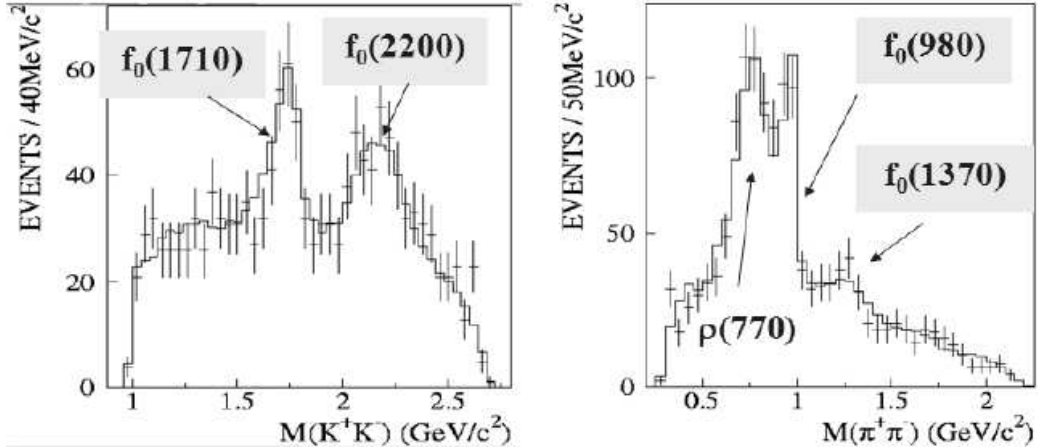}
\caption{Invariant mass spectra of $K^+K^-$ and $\pi^+\pi^-$ for the
$\chi_{c0}\to\pi^+\pi^- K^+ K^-$ decays~\cite{BESchi2pi2K}. }
\label{fig:chi0-2K2pi}
\end{figure}

For the scalar resonance production, we have studied the
$\chi_{c0}\to\pi^+\pi^- K^+ K^-$~\cite{BESchi2pi2K} and
$\pi^+\pi^-\pi^+\pi^-$~\cite{BESchi4pi} decays. Partial wave
analysis has been performed for the $\chi_{c0}\to\pi^+\pi^- K^+ K^-$
decays. The largest meson pair production rates are found for
$K^*_0(1430)\bar K^*_0(1430)$, $K^*_0(1430)\bar K^*_2(1430)+c.c.$,
and $K^*(892)\bar K^*(892)$. So it seems again that the
non-glueballs are favorably produced. The invariant mass spectra of
$K^+K^-$ and $\pi^+\pi^-$ for the $\chi_{c0}\to\pi^+\pi^- K^+ K^-$
decays are shown in Fig.\ref{fig:chi0-2K2pi}. Peaks due to
$f_0(1710)$ and $f_0(2200)$ in $K^+K^-$ invariant mass spectrum and
$\rho$, $f_0(980)$ and $f_0(1370)$ in $\pi\pi$ invariant mass
spectrum are clearly visible. For the $\chi_{c0}\to\pi^+\pi^-\pi^+
\pi^-$, the peaks due to $\rho$, $f_0(980)$ and
$f_0(1370)/f_2(1270)$ are also visible in the invariant $\pi^+\pi^-$
mass spectrum~\cite{BESchi4pi} as shown in Fig.\ref{fig:chi-4pi}(a).
For the $\chi_{c2}\to\pi^+\pi^-\pi^+ \pi^-$, only the $\rho$ peak is
predominant as shown in Fig.\ref{fig:chi-4pi}(b) without any obvious
information on $f_0$ resonances.

\begin{figure}[htbp]
\includegraphics[width=14cm,height=6cm]{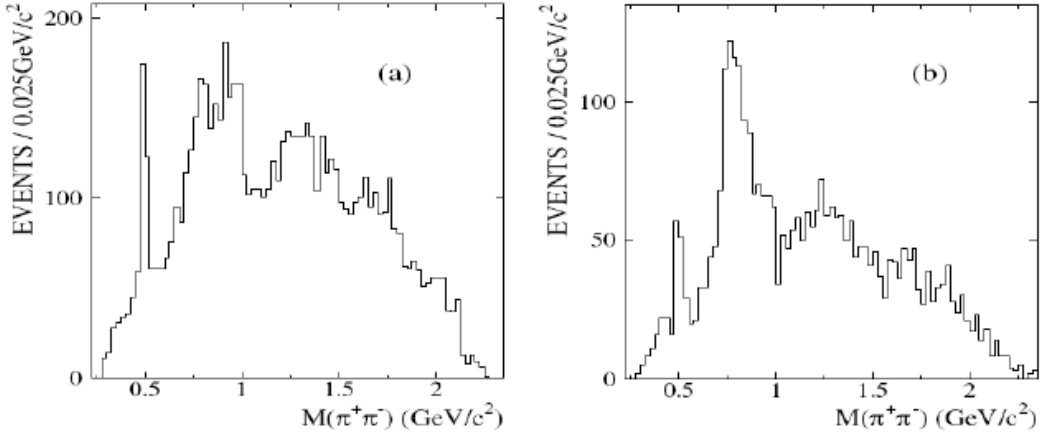}
\caption{Projections of $\pi^+\pi^-$ invariant mass under the (a)
$\chi_{c0}$ and (b) $\chi_{c2}$ peaks (two entries per event) for
$\psi^\prime\to\pi^+\pi^-\pi^+ \pi^-$~\cite{BESchi4pi}. }
\label{fig:chi-4pi}
\end{figure}

\section{Summary}

The $\psi$ radiative decays are expected to be the best place to
look for glueballs. Three $0^{++}$ peaks are observed around
$f_0(1500)$, $f_0(1710)/f_0(1790)$ and $f_0(2020)/f_0(2100)$ in
various $\psi$ radiative decays.

Both $\psi$ hadronic decays against $\phi/\omega$ and $\chi_{c0}$
decays seem favoring the non-glueball production. For $\psi$
hadronic decays against $\phi/\omega$, the $\sigma/f_0(600)$,
$f_0(980)$, $f_0(1370)$, $f_0(1710)$ and $f_0(1790)$resonances are
clearly observed. For the $\chi_{c0}\to\pi^+\pi^- K^+ K^-$ and
$\pi^+\pi^-\pi^+\pi^-$ decays, the $f_0(980)$, $f_0(1370)$,
$f_0(1710)/f_0(1790)$, and $f_0(2200)$ resonances are clearly seen.

The fact that the $f_0(1500)$ is produced strongly only in the
glueball favorable $\psi$ radiative decays and weaker than other
$f_0$ resonances in the non-glueball favorable reactions deserves
further attention. To study $f_0\to\eta\eta$, $\eta\eta'$,
$\gamma\phi$, $\gamma\rho$, $\gamma\omega$ from $\psi$ radiative
decays at BESIII will shed important light on the nature of these
$0^{++}$ resonances~\cite{close}.

\begin{theacknowledgments}
The BES collaboration thanks the staff of BEPC and computing center
for their hard efforts. This work is supported in part by the
National Natural Science Foundation of China under contracts Nos.
10491300, 10225524, 10225525, 10425523, 10625524, 10521003, the
Chinese Academy of Sciences under contract No. KJ 95T-03, the 100
Talents Program of CAS under Contract Nos. U-11, U-24, U-25, and the
Knowledge Innovation Project of CAS under Contract No. U-602 (IHEP),
the National Natural Science Foundation of China under Contract No.
10225522 (Tsinghua University), and the Department of Energy under
Contract No. DE-FG02-04ER41291 (U. Hawaii).
\end{theacknowledgments}



\begin{thebibliography}{9}

\bibitem{PDG} Particle Data Group, J. Phys. G33 (2006) 1.

\bibitem{BESg4pi}
J.~Z.~Bai et al. (BES Collaboratin),  Phys. Lett. B472 (2000) 207.

\bibitem{BESg2pi}
M.~Ablikim et al. (BES Collaboration), Phys. Lett. B 642 (2006) 441.

\bibitem{BESg2K}
J.~Z.~Bai et al. (BES Collaboration), Phys.Rev.D68 (2003) 052003.

\bibitem{BESgomegaphi}
M.~Ablikim et al. (BES Collaboration), Phys. Rev. Lett. 96 (2006)
162002.

\bibitem{Mark3} D.~V.~Bugg et al., Phys. Lett. B353 (1995) 378.

\bibitem{BESg2pi(2S)}
M.~Ablikim et al. (BES Collaboration), arXiv: 0710.2324 [hep-ex].

\bibitem{X1810} B.~A.~Li, Phys. Rev. D74 (2006) 054017; P.~Bicudo et
al., Eur. Phys. J. C52 (2007) 363; K.~T.~Chao, hep-ph/0602190;
D.~V.~Bugg, hep-ph/0603018; X.~G.~He et al., Phys. Rev. D73 (2006)
114026; Q.~Zhao et al., Phys. Rev. D74 (2006) 114025; J.~Rosner,
Phys. Rev. D74 (2006) 076006.

\bibitem{BESomega2pi}
M.~Ablikim et al. (BES Collaboration), Phys.Lett. B598 (2004) 149.

\bibitem{BESomega2K}
M.~Ablikim et al. (BES Collaboration), Phys. Lett. B603 (2004) 138.

\bibitem{BESphi2pi}
M.~Ablikim et al. (BES Collaboration), Phys. Lett. B607 (2005) 243.

\bibitem{zou1} B.~S.~Zou and D.~V.~Bugg, Phys. Rev. D48 (1993) 3948.

\bibitem{zheng} H.~Q.~Zheng et al., Nucl. Phys. A733 (2004) 235.

\bibitem{BESpsi2pi}
M.~Ablikim et al. (BES Collaboration), Phys. Lett. B645 (2007) 19.

\bibitem{f0(980)}  E.~van~Beveren et al., Phys. Lett. B641 (2006)
265; L.Roca et al., Nucl. Phys. A744 (2004) 127; G.~Janssen et al.,
Phys. Rev. D52 (1995) 2690.

\bibitem{zhao2} Q.~Zhao, Phys. Rev. D72 (2005) 074001.

\bibitem{BESchi2pi2K}
M.~Ablikim et al. (BES Collaboration), Phys. ReV. D72 (2005) 092002.

\bibitem{BESchi4pi}
M.~Ablikim et al. (BES Collaboration), Phys. ReV. D70 (2004) 092002.

\bibitem{close} F.~E.~Close et al., Phys.Rev. D67 (2003) 074031;
H.~Nagahiro et al., arXiv:0803.4460 [hep-ph].

\end{thebibliography}
\end{document}